\documentclass[iop,apjl]{emulateapj}

\usepackage{color,natbib}

\newcommand{\nraoack}{The National Radio Astronomy Observatory is a
facility of the National Science Foundation operated under cooperative
agreement by Associated Universities, Inc.}

\newcommand{\hi}{\ion{H}{1}}
\newcommand{\vlsr}{\ensuremath{V_{LSR}}}
\newcommand{\Msol}{\ensuremath{\mathrm{M_{\odot}}}}
\newcommand{\kms}{\ensuremath{\mathrm{km\,s^{-1}}}}

\newcommand{\Lsol}{\ensuremath{\mathrm{L_{\odot}}}}
\newcommand{\cmc}{\ensuremath{\mathrm{cm^{-3}}}}
\newcommand{\mlim}{\ensuremath{M_{HI}^{lim}}}
\newcommand{\mhi}{\ensuremath{M_{HI}}}
\newcommand{\hilv}{\ensuremath{M_{HI}^{lim}/L_V}}
\newcommand{\hilvd}{\ensuremath{M_{HI}/L_V}}
\newcommand{\hims}{\ensuremath{M_{HI}^{lim}/M_*}}
\newcommand{\hidyn}{\ensuremath{M_{HI}^{lim}/M_{dyn}}}
\newcommand{\hidynd}{\ensuremath{M_{HI}/M_{dyn}}}
\newcommand{\mdyn}{\ensuremath{M_{dyn}}}
\newcommand{\MLsol}{\ensuremath{\mathrm{M_{\odot} / L_{\odot}}}}
\newcommand{\sig}{\ensuremath{\sigma_{15}}}
\newcommand{\dmw}{\ensuremath{D_{MW}}}
\newcommand{\rvir}{\ensuremath{R_{vir}}}
\newcommand{\gass}{{\sc GASS}}
\newcommand{\alf}{{\sc ALFALFA}}
\newcommand{\hipass}{{\sc HIPASS}}
\newcommand{\gbt}{{\sc GBT}}
\newcommand{\lab}{{\sc LAB}}
\newcommand{\gbtidl}{{\sc GBTIDL}}

\slugcomment{DRAFT: \today}

\shorttitle{\hi\  Limits for Galactic dSphs}
\shortauthors{Spekkens et al.}

\begin{document}


\title{The Dearth of Neutral Hydrogen in Galactic Dwarf Spheroidal Galaxies}

\author{Kristine Spekkens and Natasha Urbancic}

\affil{Royal Military College of Canada, Department of Physics, PO Box
17000, Station Forces, Kingston, Ontario, Canada K7K 7B4}

\email{kristine.spekkens@rmc.ca}

\author{Brian S. Mason} 

\affil{National Radio Astronomy Observatory, 520 Edgemont Road,
Charlottesville, VA 22903-2475}



\author{Beth Willman}

\affil{Haverford College, 370 Lancaster Avenue, Haverford, PA 19041, USA}

\author{James E. Aguirre}

\affil{University of Pennsylvania, Department of Physics and
Astronomy, 209 South 33rd Street, Philadelphia, PA 19104}

\begin{abstract}
We present new upper limits on the neutral hydrogen (\hi) content within the stellar half-light ellipses of 15 Galactic dwarf spheroidal galaxies (dSphs), derived from pointed observations with the Green Bank Telescope (GBT) as well as Arecibo L-band Fast ALFA (\alf) survey and Galactic All-Sky Survey (\gass) data. All of the limits \mlim\ are more stringent than previously reported values, and those from the GBT improve upon contraints in the literature by a median factor of 23. Normalizing by $V$-band luminosity $L_V$ and dynamical mass $M_{dyn}$, we find $\hilv \sim 10^{-3} \,\MLsol$ and $\hidyn \sim 5\times10^{-5}$, irrespective of location in the Galactic halo.  Comparing these relative \hi\ contents to those of the Local Group and nearby neighbor dwarfs compiled by McConnachie, we find that the Galactic dSphs are extremely gas-poor. Our \hi\ upper limits therefore provide the clearest picture yet of the environmental dependence of the \hi\ content in Local Volume dwarfs.  If ram pressure stripping explains the dearth of \hi\ in these systems, then orbits in a relatively massive Milky Way are favored for the outer halo dSph Leo~I, while Leo~II and Canes~Venatici~I have had a pericentric passage in the past. For Draco and Ursa~Minor, the interstellar medium mass that should accumulate through stellar mass loss in between pericentric passages exceeds \mlim\ by a factor of $\sim30$. In Ursa~Minor, this implies that either this material is not in the atomic phase, or that another mechanism clears the recycled gas on shorter timescales.

 \end{abstract}


\section{Introduction}
\label{intro}

\begin{deluxetable*}{lccccccccccc}
\tablecaption{New \hi\ Limits for dSphs \label{tab}}
\tablehead{ \colhead{Name}  & \colhead{$D_{\odot}^a$}  & \colhead{\vlsr$^a$} &  \colhead{$L_V^a$}   & \colhead{$r_h^a$}  & \colhead{$\epsilon^a$}  & \colhead{$\sigma_*^a$}  & \colhead{\hi}   & \colhead{\sig}  & \colhead{\mlim} & \colhead{\hilv} & \colhead{\hidyn}  \\
                                              & \colhead{(kpc)}         &  \colhead{(\kms)}  &\colhead{(\Lsol)} & \colhead{(arcmin)}   &  & \colhead{(\kms)}   &                        \colhead{Source}   &     \colhead{(mJy/beam)}           &     \colhead{(\Msol)}&   \colhead{(\MLsol)}    &      \\
                        \colhead{(1)}       & \colhead{(2)}& \colhead{(3)} & \colhead{(4)} &  \colhead{(5)} & \colhead{(6)}  & \colhead{(7)} & \colhead{(8)} & \colhead{(9)} & \colhead{(10)} & \colhead{(11)}  & \colhead{(12)} }            
\startdata
Milky Way:\\
Segue I                &   23   &   203 &   3.4E2 &  4.4  & 0.48 & 3.9   &\gbt\     &     1.1   &    11     &   3.1E-2        &    4.1E-5                \\
Sagittarius dSph  &   26   &  149  &   2.2E7 &  340   & 0.64 & 11.4   & \gass\  &  31   &   7880    &  3.7E-4    &    4.1E-5                 \\
Ursa Major II        &   32   &  -113 &   4.1E4 &  16    & 0.63 &  6.7    & \gbt    &    2.0  &    74     &   1.8E-3    &     1.9E-5                 \\
Bootes II              &   42    & -107 &   1.0E3 &  4.2   & 0.21 & 10.5    & \gbt      &  1.2  &    38     &   3.7E-2    &     1.2E-5                 \\
Coma Berenices  &  44    &  104  &   3.7E3 &  6.0   & 0.38 &  4.6     & \gbt      &  1.8  &    62     &   1.7E-2    &     6.5E-5                 \\
Bootes III             &   47   &   210  &   1.8E4 & 48$^b$  & 0.50$^b$ & 14.0    & \alf    &    2.0   &   1080    &    6.1E-2     &    1.5E-5    \\  
Bootes I               &   66   &   110   &  2.8E4 & 13     & 0.39 & 2.4      & \gbt    &    1.6  &  252   &    8.9E-3     &      3.1E-4                \\
Draco                  &   76   &   -274  &  2.8E5 &  10    & 0.31 & 9.1      & \gbt    &  0.75  &  133     &    4.7E-4    &     1.2E-5                 \\
Ursa Minor           &  76   &    -233 &  2.8E5 &  8.2   & 0.56 & 9.5      &\gbt     &  0.61  &   63     &     2.2E-4   &     6.6E-6                  \\ 
Sextans               &  86    &    216  &  4.5E5 &  28    & 0.35 & 7.9      & \gass &   29  &  9430  &     2.1E-2          &  3.8E-4                     \\
Carina                 &   105  &   204  &  3.7E5 &  8.2   & 0.33 & 6.6       &\gass &    24  &  4780  &    1.3E-2         &    7.6E-4               \\
Leo V                   &   178  & 171   &  1.0E4 & 2.6    & 0.50 & 3.7      & \gbt    &    0.72 &  403    &     3.9E-2          &    3.7E-4                  \\
Canes Venatici I  &   218  &  42    &  2.4E5 & 8.9    & 0.39 & 7.6       & \gbt    &    0.98  &  1170 &      4.9E-3          &    6.1E-5                \\
Leo II                   &   233  &  78    &  7.1E5 & 2.6    & 0.13 & 6.6       & \alf     &     2.0  &  1960  &     2.8E-3          &    4.3E-4                \\
Leo I                    &   254  &  277  &  5.4E6 & 3.4     & 0.21 & 9.2       & \alf     &     2.2  &   3560 &     6.6E-4          &    3.0E-4                 \\

\hline
Local Group: &&&&&&&     \\ 
Cetus                    & 755   & -88   & 2.8E6 & 3.2 &    0.33     &  8.3 & GASS & 23 & 2.3E5 & 8.3E-2 &  4.4E-3\\
Tucana                 & 887   & 188   & 5.6E5 &  1.1 & 0.48       & 15.8 & GASS & 21 & 2.9E5 & 5.2E-1 & 3.5E-3    \\
Andromeda XVIII  & 1213 & -325 & 5.0E5 & 0.92 & \nodata$^c$ & 9.7 & LAB & 183 &  4.8E6 & 9.5E0  &  2.7E-1  
\enddata
\tablecomments{ Col.\ 1: dSph name;  Col.\ 2: distance from the Sun; Col.\ 3: Systemic velocity \vlsr\  of the stellar component, in the LSRK frame; Col.\ 4: $V$-band luminosity, corrected for Galactic extinction; Col.\ 5: elliptical half-light radius $r_h$;  Col.\ 6: ellipticity $\epsilon$ of the stellar component; Col.\ 7: Velocity dispersion of the stellar component; Col.\ 8: source of the single-dish \hi\ spectrum: ALFALFA, GASS, LAB or our own GBT data;  Col.\ 9: RMS noise of the \hi\ spectrum at $15\,\kms$ resolution, at \vlsr; Col.\ 10: 5\sig\ upper limit on the \hi\ mass; Col.\ 11: Upper limit on the ratio of \hi\ mass to $L_V$; Col.\ 12: Upper limit on the ratio of \hi\ mass to dynamical mass \mdyn . \\
$a$: taken from M12 unless otherwise noted. \\
$b$: approximate values from \citet{grillmair09}.\\
$c$: no estimate available. Because $r_h<< \theta$ for the LAB survey, $N_b=1$ in eq.~\ref{eq:hi} irrespective of the value of $\epsilon$.}
\end{deluxetable*}

The origin and evolution of the Milky Way's dwarf galaxy satellites address several problems in cosmological galaxy formation \citep[e.g.][]{weinberg13}. In particular, their gas content provides an important clue to their evolutionary histories: it has long been appreciated that dwarf spheroidal galaxies (dSphs), preferentially located within the virial radius of the Milky Way, tend to be deficient in neutral hydrogen (\hi) compared to the dwarf irregulars located farther out \citep[e.g.][]{einasto74,lin83}. A variety of searches for \hi\ in the Galactic dSphs have been performed over the years \citep[e.g.][hereafter GP09]{knapp78,blitz00,bouchard06,grcevich09}. Although some detections have been claimed \citep[e.g.][]{carignan98,blitz00,bouchard06}, most were not confirmed by deeper observations (see the discussion in GP09). These non-detections suggest that environmental processes play a key role in shaping the morphologies of Galactic dSphs. Stringent upper limits on the \hi\ content of the latter may elucidate the details of these mechanisms, as well as the basic Milky Way and dSph properties on which they depend.

The \hi\ content of nearby dwarf galaxies has been most recently examined by GP09, who used non-detections in reprocessed \hi\ Parkes All-Sky Survey \citep[\hipass,][]{putman03} data and in Leiden-Argentine-Bonn \citep[\lab,][]{kalberla05} survey data  to constrain Galactic accretion rates due to satellite infall. However, the sensitivities of those surveys\footnote{Errors in the \hi\ mass equations in GP09 imply that the upper limits they derive from \hipass\ and \lab\ data are under-estimated by factors of $\sim20$ and $\sim15$, respectively.} produce relatively weak upper limits on the faintest nearby dSphs, while those for dSphs  in the outer halo don't strongly exclude characteristic values of gas-rich Local Volume dwarfs farther out. Pointed observations with large single dishes such as the Robert C.\ Byrd Green Bank Telescope (GBT) have the potential to search for \hi\ down to much lower levels, and higher sensitivity and spectral resolution surveys such as the Arecibo Legacy Fast ALFA Survey \citep[\alf,][]{giov05} and the Galactic All-Sky Survey \citep[\gass,][]{mcclure09}, respectively, are also now available. These developments warrant a re-examination of the \hi\ content of the Galactic dSphs, which we carry out here for the subset of that population that are well-separated in velocity from the Milky Way's \hi\ disk and high-velocity clouds (HVCs).

\section{Sample Selection and Observations}
\label{data}

We examine the \hi\ content of all dSphs within $300\,$kpc  of the Galactic Center that have:
\begin{enumerate}
 \item distances $D$ and radial velocities \vlsr\ measured from resolved stellar population studies;
 \item a value of \vlsr\ that does not fall in the velocity range of either Galactic disk or HVC \hi\ emission along the dSph line-of-sight in the deepest available data. 
 \end{enumerate}
 Comparing to the Milky-Way subgroup compilation of \citet[][hereafter M12]{mcconnachie12}, the first criterion omits Pisces II from our sample for its lack of a measured \vlsr . The second criterion provides a ``clean" sample in which to search for \hi\ down to low levels. It excludes Canis Major, Segue~II, Hercules, Willman~I, and Ursa Major~I because \vlsr\ overlaps with the Galactic \hi\ disk, as well as Fornax, Sculptor and Leo~IV because \vlsr\ falls in the velocity range of known HVC complexes (G09, \citealt{moss13}). We find faint high-velocity gas at $-200 \, \kms < \vlsr\ < -100 \, \kms$  in our GBT spectra of Canes Venatici~I (CVn~I) and Canes Venatici~II;  \vlsr\ of the latter dSph falls in this velocity range, and we therefore omit it from the sample. We discuss the potential impact of our selection criteria on our conclusions in \S\ref{discuss}.
 
 The basic properties of the 15 Galactic dSphs that satisfy our selection criteria are given in Table~\ref{tab}. Unless otherwise noted, we use the measured stellar properties compiled by M12 (2013 May 30 database version) for all systems.
  
We examine the \hi\ environment of each sample dSph using \gass\ or \alf\ survey data where available, or using \lab\ data otherwise. Assuming that any putative \hi\ in these systems is in dynamical equilibrium, we expect the width of this distribution to resemble that of the stars. To maximize our sensitivity to this \hi\ signal, we therefore smooth all spectra to a resolution of 15\,\kms, which corresponds to the characteristic stellar velocity distribution width in the sample dSphs (see Table~\ref{tab}). We examine the most sensitive survey data available at \vlsr\ near the stellar centroid of each dSph, and find no \hi\ emission. Since the \gass\ and \lab\ surveys also afford new \hi\ limits on the Local Group dSphs Cetus, Tucana and Andromeda~XVIII, we include them at the bottom of Table~\ref{tab}. 


We obtained deep, pointed GBT \hi\ observations  for a subset of the sample in 2014 May and 2014 June (program AGBT14A\_284). We used the GBT Spectrometer with a bandpass of $12.5\,$MHz and $1.56\,$kHz channels centered at $V_{LSRK}=0$ to obtain an ``on" spectrum along the line-of-sight to each dSph, as well as an off-target reference spectrum of equal integration time to subtract from the ``on" scan and flatten the spectral baseline. For dSphs with half-light diameters $2r_h$ that exceed the 9.1\arcmin\ full-width at half maximum $\theta$ of the GBT beam, additional offset pointings of comparable depth were obtained to fully sample the optical half-light ellipse $\pi (1-\epsilon)r_h^2$, where $\epsilon$ is the ellipticity of the stellar distribution. The data were reduced using the standard \gbtidl\footnote{ \tt http://gbtidl.nrao.edu/} routine {\it getps} and smoothed to a resolution of $15\,$\kms. 

We find no \hi\ emission at \vlsr\ of the dSphs targeted by our GBT observations. The source of the most sensitive $15\,$\kms--resolution \hi\ spectrum at the location of each sample dSph, and the root mean square (RMS) noise \sig\ per beam of that spectrum at \vlsr, are given in cols.~8~and~9 of Table~\ref{tab}, respectively.

\section{Calculating \hi\ Limits}
\label{limits}

\begin{figure*}
\epsscale{1.0}
\plotone{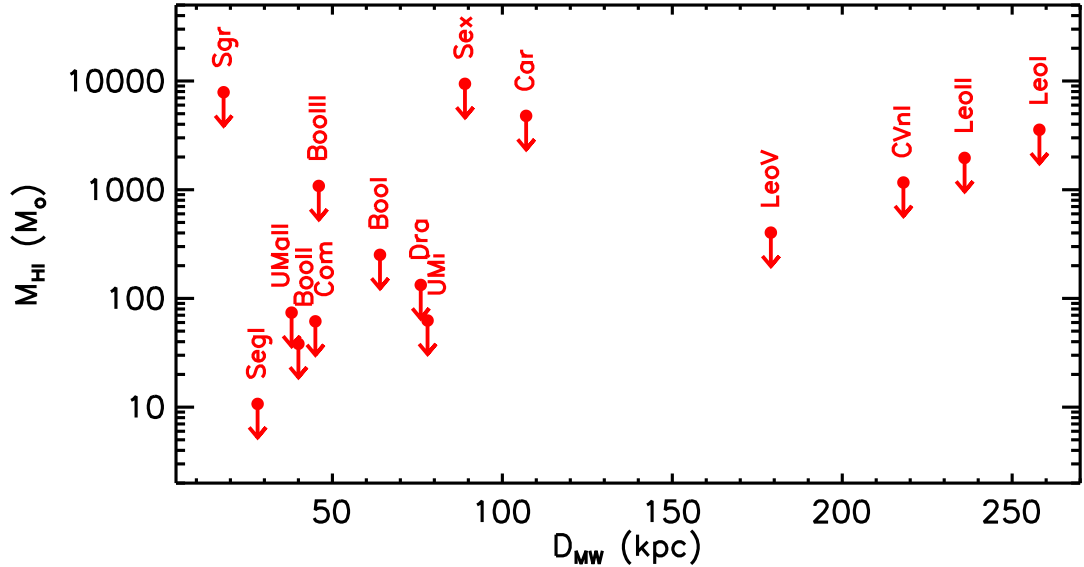}
\caption{Upper limits on the \hi\ masses of Galactic dSphs within their half-light ellipses, plotted as a function of Galactocentric distance.}
\label{fig:hi}
\end{figure*}

We use the non-detection at \vlsr\ of each dSph to compute upper limits on the \hi\ mass within its half-light ellipse. The \hi\ mass \mhi\ of a detected source is given by:
\begin{equation}
\mhi = 2.36 \times 10^{-4} D^2 \int{S dV} \,\,\,\,\Msol ,
\end{equation}
where $D$ is the distance to the source in kpc and $\int{S\,dV}$ is the integrated \hi\ flux across the source in $\rm{mJy\,\kms}$. We compute $5\sigma$, $15\,\kms$--width upper limits on the \hi\ mass within the half-light ellipse of each dSph:  
\begin{equation}
\mlim = 0.0177 \, D^2 \, \sig \sqrt{N_b} \,\,\,\,\Msol ,
\label{eq:hi}
\end{equation}
where \sig\ is the RMS noise in mJy/beam of the $15\,\kms$--resolution spectrum at \vlsr. In eq.~\ref{eq:hi}, $N_b$ is the number of beams across the source, rounded up to the nearest integer:
\begin{equation}
N_b =  \left \lceil \frac{2.77(1-\epsilon)r_h^2}{\theta^2} \right \rceil  \,\,\,,
\end{equation}
where $\theta=9.1$\arcmin\ for our GBT spectra, $\theta = 4.0$\arcmin\ for \alf\ \citep{giov05}, $\theta = 16$\arcmin\ for \gass\ \citep{kalberla10}, and $\theta = 40$\arcmin\ for \lab\ \citep{kalberla05}.
The values of \mlim\ for the sample dSphs are given in col.~10 of Table~\ref{tab}, and are plotted in Fig.~\ref{fig:hi} as a function of Galactocentric distance \dmw.

\section{Results: new \hi\ Limits for Galactic dSphs}
\label{results}

The \hi\ limits that we calculate for Galactic dSphs are in the range $11\,\Msol < \mlim\ < 9430\,\Msol$ (Table~\ref{tab}); all of them are more stringent than previously reported values. In particular, \mlim\ derived from our GBT observations yield tighter upper limits than corresponding values in the literature by factors ranging from 2.6 \citep[][for Draco]{knapp78} to 610 (G09, for CVn~I, after correcting for errors in their limiting \hi\ mass relations), with a median improvement of a factor of 23. 

\begin{figure*}
\epsscale{1.0}
\plotone{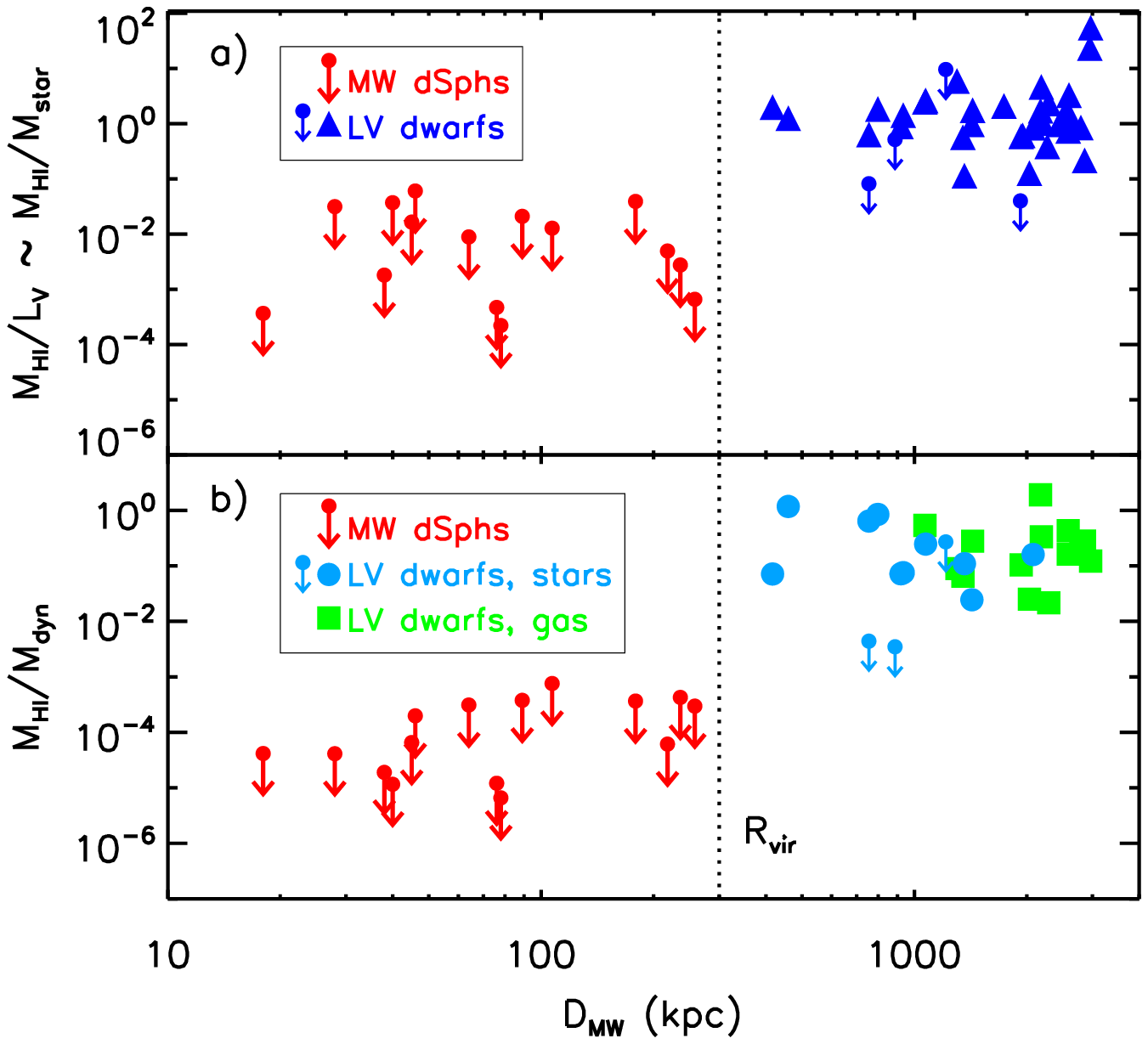}
\caption{\hi\ content of the Milky Way dSphs and Local Volume dwarfs, normalized by $a)$ $V$-band luminosity $L_V$ ($\sim\,$ the stellar mass $M_*$), and $b)$ dynamical mass \mdyn. In $a)$, the red arrows show \hilv\ for the sample Milky Way dSphs, and the blue filled triangles and arrows show \hilvd\ and \hilv, respectively, for systems classified as Local Group satellites or nearby neighbours by M12.  In $b)$, the red arrows show \hidyn\ for the sample Milky Way dSphs, where \mdyn\ is computed from stellar kinematics. The light blue filled circles and arrows show \hidynd\ and \hidyn\ for Local Volume satellites, respectively, where \mdyn\ is computed from stellar kinematics. The green filled squares show \hidynd\ for Local Volume satellites where gas kinematics are used to compute \mdyn . The vertical dotted line in both panels shows the approximate virial radius of the Milky Way, $R_{vir}=300\,$kpc.}
\label{fig:hilv}
\end{figure*}

Fig.~\ref{fig:hilv} shows \mlim\ normalized by $a)$ $V$-band luminosity $L_V$, and $b)$ dynamical mass \mdyn\ within the half-light radius, as a function of \dmw.  We use the relation of \citet{walker09c} to compute \mdyn\ from the stellar kinematics of each dSph, as compiled by M12:
\begin{equation}
\mdyn = 580 \, r_h \, \sigma_*^2  \,\,\,\Msol ,
\label{eq:dyn}
\end{equation}
where $\sigma_*$ is the stellar velocity dispersion in \kms\  and $r_h$ is in parsecs.

Since $M_*/L_V \sim 1\,\MLsol$ for the old, metal poor stellar populations characteristic of many dSphs \citep[e.g.][]{martin08}, Fig.~\ref{fig:hilv}$a$ effectively plots \hims. Uncertainties in the initial mass function and stellar evolution of the faintest dwarfs should move the points by less than $\sim0.5\,$dex \citep{martin08,geha13}.   
 
To compare the \hi\ content of the sample dSphs to that of more isolated nearby systems, we also plot \hilvd\ and \hidynd\ for Local Volume dwarfs in Fig.~\ref{fig:hilv}. We consider all dwarfs classified as Local Group satellites or nearby neighbors in M12 as our reference, omitting only Phoenix because the \hi\ detected near it is offset from the stellar distribution by $\gtrsim 0.5 r_h$ \citep{young07}. The dotted vertical line in Fig.~\ref{fig:hilv} shows the approximate virial radius of the Milky Way, $R_{vir} = 300\,$kpc \citep[e.g.][]{klypin02}.

To compute \hilvd\ for the Local Volume dwarfs, we take the \hi\ detection and luminosity information directly from M12. Only 4/37 systems are not detected in \hi: we plot \hilv\ for them using our new limits for Cetus, Andromeda XVIII and Tucana (Table~\ref{tab}), and the 5$\sigma$ upper limit $\mlim = 8 \times 10^4\,\Msol$ for KKR~25 obtained by \citet{begum05}. 

We also plot \hidynd\ or \hidyn\ for the 27/37 Local Volume dwarfs where stellar or \hi\ kinematics have been measured. We use stellar kinematics where available, and use the \hi\ kinematics otherwise; for systems with detected rotation $V_r$, we substitute $\sigma_*^2 \rightarrow \sigma^2 + V_r^2$ in eq.~\ref{eq:dyn}. More sophisticated estimates of \mdyn\ could be obtained by examining the kinematics of each system in detail, but are unlikely to differ from those in Fig.~\ref{fig:hilv}$b$ by more than $\sim0.5\,$dex. 

Fig.~\ref{fig:hilv} shows that we find $\hilv \sim 10^{-3} \,\MLsol$ and $\hidyn \sim 5\times10^{-5}$ for the dSphs, for all $\dmw < R_{vir}$. Our upper limits therefore conclusively demonstrate that if there is any \hi\ in these systems, it is negligible compared to both their stellar and dynamical masses.  By contrast, most Local Volume dwarfs have relative \hi\ contents $\hilvd \sim 1 \,\MLsol$ and $\hidynd \sim 0.1$. Our values of \mlim\ for the sample dSphs therefore imply that they are extremely gas-poor relative to the bulk of the Local Volume dwarfs.

\section{Discussion}
\label{discuss}

The environmental dependence of the \hi\ content of Local Group galaxies is well-known (see \S\ref{intro}), but Fig.~\ref{fig:hilv} presents the clearest picture of this phenomenon yet. Our data show for the first time that, notwithstanding the Magellanic System \citep{bruns05}, Cetus, Tucana and KKR~25, the transition between gas-rich and gas-poor dwarfs in the vicinity of the Milky Way is abrupt and located near \rvir .
Even our weakest upper limits imply that \hilvd\ and \hidynd\ for the sample dSphs are at least $\sim100$ and $\sim1000$ times lower than the characteristic value for Local Volume dwarfs, irrespective of their distance from the Galactic center. 

For this study, we selected only the 15/26 Galactic dSphs with \vlsr\ that do not overlap with the Milky Way's \hi\ disk or HVC emission (see \S\ref{data}). This produces a ``clean" sample from an \hi\ search perspective, but also raises the possibility that the excluded dSphs contain gas in excess of the \hilv\ and \hidyn\ reported here. Our cursory examination of the \gass\ and \lab\ data for these objects indicates that this is not the case. G09 reached a similar conclusion using \hipass\ data, classifying \hi\ clouds claimed to be associated with Sculptor \citep{carignan98} and Fornax \citep{bouchard06} as ambiguous detections. None of these clouds lie within the half-light ellipse of the dSph in its vicinity. 

dSphs at a given \dmw\ exhibit a range of \vlsr\ because they are bound to the Milky Way (e.g.\ M12), and the latter's \hi\ disk and HVCs fill a sizeable fraction of the spatial and spectral search volume \citep[e.g.][]{peek11,moss13}: an overlap in \vlsr\ between Galactic \hi\ clouds and dSph stars is therefore unlikely to imply a physical association between them.  An investigation of the \hi\ content of dSphs in these confused fields requires deep, wide-field \hi\ mapping, and even then associating \hi\ features with stellar systems at the same location is not straightforward  \citep[e.g.][]{carignan98,bouchard06,grcevich09}. We defer a detailed examination of the \hi\ content of the dSphs confused with Galactic emission to a future publication (Spekkens et al.\ 2015, in preparation), but argue that the dearth of \hi\ in the sample presented here is likely representative of the Galactic dSph population as a whole. 


Ram pressure stripping is thought to explain the absence of \hi\ in the Galactic dSphs relative to Local Volume dwarfs (e.g.\ \citealt{lin83}, \citealt{blitz00}, G09), though tidal effects \citep[e.g.][]{read06,mayer06}, resonances \citep[e.g.][]{donghia09} and feedback from star formation \citep[e.g.][]{gatto13} may also play a role. \citet{gatto13} show that numerical simulations are needed to accurately model gas stripping. Nonetheless, a scaling of the analytic \citet{gunn72} criterion reasonably describes the conditions under which ram pressure stripping occurs in dSph progenitors:
\begin{equation}
\rho_{MW} v_{orb}^2 \gtrsim 5 \rho_{d} \sigma_*^2 \,\,\,,
\label{rp}
\end{equation}
where $\rho_{MW}$ and $\rho_{d}$ are the densities of the hot halo gas and dSph interstellar medium, and $v_{orb}$ is the dSph orbital velocity. The small \mlim\ for the outer halo dSphs CVn~I, Leo~II and Leo~I reported here imply that this condition must be met at some point in their orbits (see also G09). 

 Adopting $\rho_{d}\sim 0.1\,$\cmc\ characteristic of Leo~T \citep{rw08}, recent estimates of the coronal density profiles of the Milky Way \citep{anderson10,gatto13,marinacci14} combined with eq.~\ref{rp} favor the pericenter distances and velocities of Leo~I implied by a Milky Way mass $M_{vir} \gtrsim 1.5\times 10^{12}\,$\Msol, at the higher end of the accepted range \citep[][]{sohn13}. Similarly, the narrow range in subhalo orbital velocities in cosmological simulations of Milky-Way sized halos \citep{bk13} suggests that, even if CVn~I and Leo~II are on their first infall, ram pressure stripping is not effective at their present \dmw. Like Leo~I, CVn~I and Leo~II have therefore likely had a pericentric passage in the past that has taken them to $\dmw \lesssim 100\,$kpc. More detailed calculations are needed to refine these arguments \citep[e.g.][]{gatto13}, but they illustrate the implications of the \mlim\ presented here for outer halo dSph orbits as well as for the basic properties of the Milky Way.  

As expected from eq.~\ref{eq:hi}, we obtain the tightest \hi\ constraints for nearby, unresolved ($N_b=1$) systems, which probe how gas is recycled into the interstellar medium (ISM) of the dSphs themselves. Assuming that a low-mass star in an old ($>10\,$Gyr) stellar population losses $\sim0.3\,\Msol$ of material after turning off the main sequence, \citet{moore11} estimate a mass loss rate of $\dot{M} \sim 5 \times 10^{-12}\,\Msol \,\mathrm{yr}^{-1} \Lsol^{-1}$.  A $L \sim 10^4\,\Lsol$ dSph with an orbital period $T \sim 1\,$Gyr therefore accumulates $\sim 50\,\Msol$ of interstellar material from stellar evolution before ram pressure sweeps away the gas at pericenter; this mass is of the same order as \mlim\ for several sample dSphs. 

Our non-detections in UMi and Draco are particularly interesting in this respect, since their $L_V$ (Table~\ref{tab}) combined with their respective $T \sim 1\,$Gyr and $T \sim5\,$Gyr inferred from proper motions \citep{lux10} imply an accumulation of interstellar material that exceeds \mlim\ by a factor of $\sim 30$. Draco is close to pericenter and thus may have just lost this gas to ram pressure, but UMi is not \citep{lux10}. As is the case in globular clusters \citep[e.g.][]{vanloon06}, \mlim\ in UMi therefore requires either that the ISM accumulated between pericentric passages is not in the atomic phase, or that another mechanism clears the recycled gas on shorter timescales.
 
\section{Conclusions}
\label{conclude}

Using a combination of pointed observations from the GBT as well as \alf\ and \gass\ survey data, we have presented new upper limits on the \hi\ content within the half-light ellipses of the 15 Galactic dSphs for which \vlsr\ does not overlap with Galactic \hi\ disk or HVC emission. Our non-detections imply values of \mlim\ that are more stringent than those available in the literature for all sample dSphs, while our GBT spectra improve upon previous limits by median factor of 23. We find upper limits on the relative \hi\ content of the dSphs to be $\hilv \sim 10^{-3} \,\MLsol$ and $\hidyn \sim 5\times10^{-5}$ for all $\dmw < \rvir$: any \hi\ present in these systems is clearly negligible compared to both their stellar and dynamical masses. We compare the relative \hi\ content of the dSph sample to that of the Local Group and nearby neighbour dwarfs compiled by M12. With few exeptions, these Local Volume dwarfs are gas-rich with $\hilvd \sim 1 \,\MLsol$ and $\hidynd \sim 0.1$, and the sample dSphs are extremely gas-poor by comparison. 

Our data show for the first time that the transition between gas-rich and gas-poor dwarfs near the Galaxy is abrupt and located near \rvir, in the clearest picture yet of the environmental dependence of their \hi\ content.  
Assuming that ram-pressure stripping is the dominant mechanism responsible for this phenomenon, we discuss how the dearth of \hi\ in outer halo dSph Leo~I favors orbits for that dwarf in a relatively massive Milky Way, while that for Leo~II and CVn~I implies that these dSphs have had a pericentric passage in the past. We also discuss how the ISM accumulated in dSphs between pericentric passages due to stellar mass loss exceeds \mlim\ for the nearby dSphs Draco and UMi by a factor of $\sim 30$. In UMi, this implies that either this material is not in the atomic phase or that another mechanism clears the recycled gas on shorter timescales. 

\acknowledgements

We thank R.~Giovanelli and M.~P.~Haynes for their help accessing the \alf\ survey data, and P.~Kalberla for his help with the \gass\ server. KS acknowledges support from the Natural Sciences and Engineering Research Council of Canada. This work was supported by NSF grant AST-1151462 to BW.
\nraoack\

{\it Facilities:} \facility{NRAO: GBT}


\end{document}